\documentclass[twocolumn,showpacs,preprintnumbers,amsmath,amssymb]{revtex4}

\usepackage{graphicx}
\usepackage{dcolumn}
\usepackage{bm}
\newcommand{\dr}{{\rm d}}

\begin{document}
\title{Observation of Exceptional Points in Electronic Circuits}
\author{T.~Stehmann}
\affiliation{Institute of Theoretical Physics, University of Stellenbosch, South Africa}
\author{W.D.~Heiss}
\affiliation{Institute of Theoretical Physics, University of Stellenbosch, South Africa}
\author{F.G.~Scholtz}
\affiliation{Institute of Theoretical Physics, University of Stellenbosch, South Africa}
\begin{abstract}
Two damped coupled oscillators have been used to demonstrate the
occurrence of exceptional points in a purely classical system. The implementation
was achieved with electronic circuits in the kHz-range. The experimental
results perfectly match the mathematical predictions at the exceptional point.
A discussion about the universal occurrence of exceptional points -- connecting 
dissipation with spatial orientation -- concludes the paper.
\end{abstract}
\maketitle

A surprising phenomenon occurring in systems described by
non-hermitian Hamiltonians has been observed in a number of
experiments:
the coalescence of two eigenmodes. If the system depends on some
interaction parameter $\lambda\, ,$ the value $\lambda_{\rm EP}$
at which the coalescence occurs is called an exceptional point
(EP) \cite{Kato}. At an EP, the eigenvalues \emph{and}
eigenvectors show branch point singularities
\cite{Kato,Hesa,BerryEP,Mondragon,kors} as functions of
$\lambda\, .$ This stands in sharp contrast to two-fold
degeneracies, where no singularity but rather a diabolic point
\cite{BerryDP} occurs. EPs have been described in laser induced
ionization of atoms \cite{Latinne}, in acoustical systems
\cite{Shuva}, and have actually been observed in
microwave cavities \cite{BrentanoEP,DemboEP,DemboCh}, in
optical properties of certain absorptive media \cite{Pancha,BerryPancha}, and
in ``crystals of light'' \cite{Oberthaler}. The broad variety of
physical systems showing EPs indicates that their occurrence is
\emph{generic}. So far EPs have been analyzed for the special case of
a complex symmetric effective Hamiltonian \cite{he99} used for the description of
dissipative wave mechanical systems such as in microwave cavities.

The physical interest in EPs is not only due to their universal occurrence in
virtually all problems of matrix diagonalization. It is in particular the
chiral character associated with the wave functions at the EP \cite{heha}. This has
been experimentally confirmed recently \cite{DemboCh} and also thoroughly
discussed in optics for anisotropic absorptive media \cite{beden}. The
fascinating point is the the wave function at the EP: not only is there -- at the
point of the two coalescing energies -- one and 
only one state vector, but -- for a complex symmetric matrix -- its form
is given by
\begin{equation} |\psi ^+_{{\rm EP}}\rangle =
\begin{pmatrix}0 \cr \vdots \cr  i \cr 
1 \cr 0 \cr \vdots \end{pmatrix} \quad {\rm or} \quad
|\psi ^-_{{\rm EP}}\rangle =\begin{pmatrix}0 \cr \vdots \cr  -i \cr 
1 \cr 0 \cr \vdots \end{pmatrix}.
\label{restr}
\end{equation}
Only one EP of the pair is accessible in the laboratory as only one is associated with
a negative sign of the imaginary part of the energy. The labels of the non-vanishing 
components in Eq.(\ref{restr}) are the labels of the two coalescing energies.
Note that there is no parameter dependence; the form of the state vector is robust.
We may associate this form with a circularly polarized wave. For a general
non-hermitian matrix the ratio of the two relevant components is no longer $\pm i$
but an arbitrary complex number that can in general be associated with an elliptic
polarisation \cite{beden}.

In this letter we report the results of a plain classical experiment where
an EP has been identified and so has the orientation that goes with it. It is
the case of two coupled damped oscillators. This has been proposed in
\cite{he03} for two mechanical oscillators. In the quoted paper a discussion
of the EPs is extended to general matrices being no longer complex symmetric.
In fact, the classical equations of motion for the momenta $p_j$ and coordinates
$q_j$ read for the two mechanical oscillators
\begin{equation}
\frac{\dr }{ \dr t}\begin{pmatrix}p_1 \cr p_2 \cr q_1 \cr q_2\end{pmatrix}={\cal M}
\begin{pmatrix}p_1 \cr p_2 \cr q_1 \cr q_2 \end{pmatrix}
 + \begin{pmatrix}c_1 \cr c_2 \cr 0 \cr 0\end{pmatrix}\exp (i\omega t)
\label{eom}
\end{equation}
with 
\begin{equation}
{\cal M}=\begin{pmatrix}-2g -2k_1 & 2g & -f-\omega_1^2 & f \cr
2g & -2g -2k_2 & f & -f-\omega_2^2 \cr
1 & 0 & 0 & 0 \cr
0 & 1 & 0 & 0  \end{pmatrix}
\label{mat}
\end{equation}
where $\omega _j - ik_j,\; j=1,2$ are essentially the damped frequencies 
without coupling 
and $f$ and $g$ are the coupling spring constant and damping of the coupling,
respectively. The driving force is assumed to be oscillatory with one single
frequency and acting on each particle with amplitude $c_j$. 
Here we are interested only in the stationary solution being the
solution of the inhomogeneous equation which reads
\begin{equation}
\begin{pmatrix}p_1 \cr p_2 \cr q_1 \cr q_2\end{pmatrix}=(i\omega -{\cal M})^{-1}
\begin{pmatrix}c_1 \cr c_2 \cr 0 \cr 0\end{pmatrix}\exp (i\omega t).
\label{sol}
\end{equation}
Resonances occur for the real values $\omega $ of the complex solutions of the
secular equation
\begin{equation}
\det |i\omega -{\cal M}|=0
\label{det}
\end{equation}
and EPs occur for the complex values $\omega $ where
\begin{equation}
{\dr \over \dr \omega}\det |i\omega -{\cal M}|=0
\label{der}
\end{equation}
is fulfilled together with Eq.(\ref{det}). Note that ${\cal M}$
is non-symmetric.

For easy implementation electronic circuits are used rather than the mechanical 
oscillators.
The momenta and coordinates are replaced by voltage and current, respectively.
The unperturbed undamped frequencies are given by $\omega _j^2=1/(L_jC_j)$
with $L$ and $C$ being the inductance and capacitance, while damping and coupling
are effected by resisters and mutual inductance. To allow for complex values
of the effective inductance, resisters have been used in parallel to the inductance.

\begin{figure}[h]
  \includegraphics*[scale=0.32]{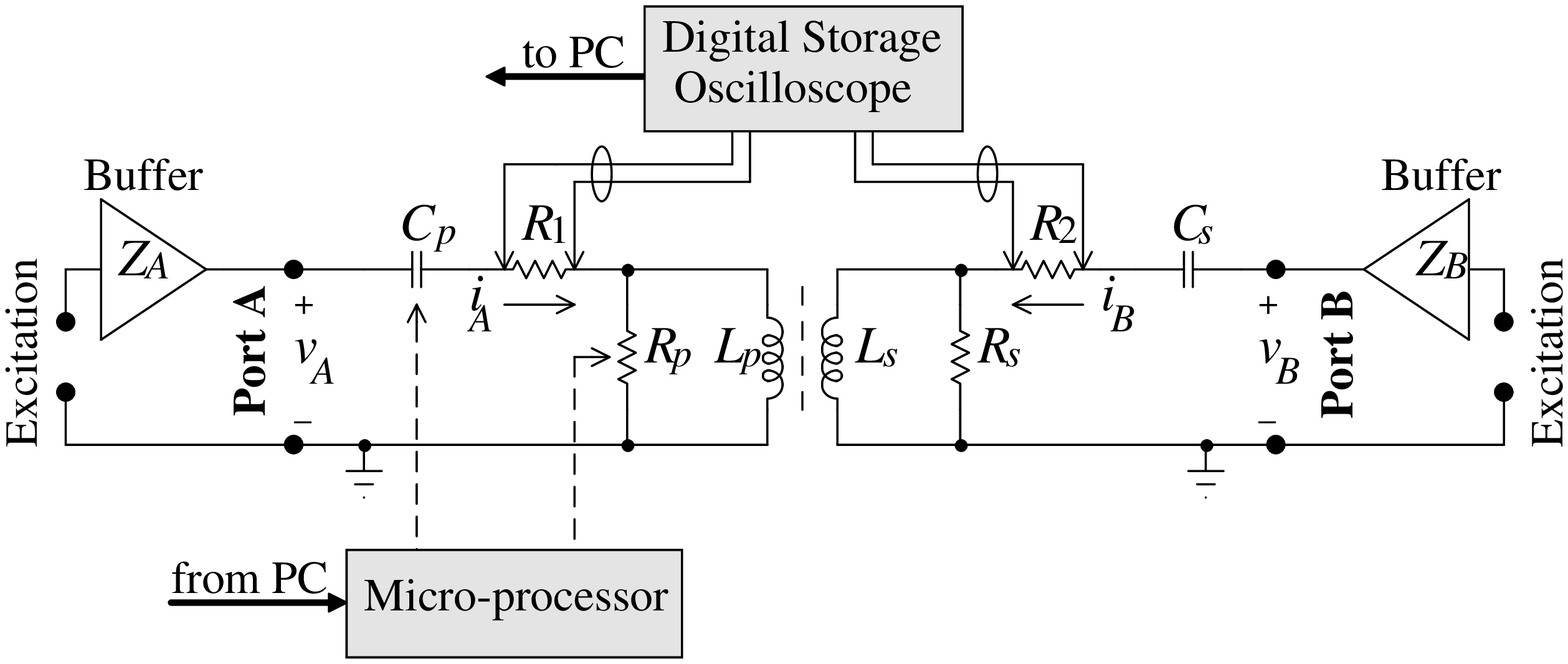}
  \caption{Measurement set-up.}
  \label{fig:setup}
\end{figure}

The actual experimental set-up is shown in Figure \ref{fig:setup}. The two circuit 
parameters that can be changed are the circuit capacitance $C_p$ and the parallel 
resistance $R_p$. Standard electronic components were used to construct the 
circuit. The component values are summarized in Table \ref{tab:components}. The 
two inductors consist of multi-layer cylindrical air core windings wound onto PVC 
plastic coil formers. The coils are placed next to each other with their geometric 
axes aligned. The distance between them can be varied in order to change the 
coupling, i.e. the mutual inductance of the two coils.
\begin{table}[h]
\caption{\label{tab:components}Component Values}
\begin{ruledtabular}
\begin{tabular}{cccc}
  Component & Value & Unit & Tolerance \\
  \hline
  $C_p\footnotemark[1]$ & 65 & nF & 5\,\%  \\
  $R_1$ &3.0 &$\Omega$ & 10\,\% \\
  $R_p\footnotemark[2]$ & 520 & $\Omega$ & 5\,\%  \\
  $L_p$ & 1.80 &mH & 5\,\%  \\
  $L_s$ & 2.55 &mH & 5\,\%  \\
  $R_s$ & 10.0 &k$\Omega$& 5\,\%  \\
  $R_2$ & 3.0 &$\Omega$ & 10\,\% \\
  $C_s$ & 47 &nF &5\,\%  \\
  $M$ & $ 0.14\sqrt{L_pL_s}=0.3$ & mH & 5\,\% \\
\end{tabular}
\end{ruledtabular}
\footnotetext[1]{Can be varied from 39.0 to 77.6\,nF in steps of 0.22\,nF.}
\footnotetext[2]{Can be varied from 430 to 610\,$\Omega$ in steps of 10\,$\Omega$.}
\end{table}
Input voltage signals $v_A$ and $v_B$ are connected to ports $A$ and $B$ of 
the circuit and are used to excite the system. The circuit response is given 
by the input currents $i_A$ and $i_B$ into ports $A$ and $B$, respectively. The 
two input currents can be determined by measuring the voltage drop over the 
series resistances $R_1$ and $R_2$. Using Ohm's Law the currents can be calculated.

The input ports $A$ and $B$ of the circuit have to be excited by ideal voltage 
sources. In a practical set-up the excitation voltage sources must have output 
impedances, $Z_A$ and $Z_B$, that are much lower than the input impedances of 
the two circuit ports. The minimum input impedance of ports $A$ and $B$ are 
approximately $R_1$ and $R_2$, respectively. Buffers are used to reduce the 
signal impedance, such that $Z_A \ll R_1$ and $Z_B \ll R_2$.


A microprocessor is used to enable automatic multiple data collection.
The microprocessor interfaces to a PC and varies the values of two circuit 
components $C_p$ and $R_p$. 
The input currents are measured with a digital oscilloscope. The oscilloscope is 
connected to the same PC as the microprocessor. The data from the oscilloscope is 
read directly into the PC where it is processed.
The complete experiment is controlled with a software package called LabVIEW. A 
program written in LabVIEW is responsible for communicating with the microprocessor 
and the digital oscilloscope and processing the measurement data.


The first step in the measurement process is to determine the system Eigenvalues. 
The Eigenvalues can be determined from the system's frequency response, i.e. the 
system's spectrum. The frequency response $F_A(\omega)$ of the system at port $A$ 
is given by \cite{nilss}

\begin{equation}
  F_{A}(\omega) = \frac{I_{A}(\omega)}{V_{A}(\omega)} =
    \frac{\displaystyle\sum_{i=0}^{M}a_i \omega^i}
         {\displaystyle\sum_{i=0}^{N}b_i \omega^i}\,\,,
\label{eq:responseA}
\end{equation}
with $M \le N$. Here $V_{A}(\omega)$ and $I_{A}(\omega)$ are the spectra of the 
applied voltage 
signal and the input current response at port $A$, respectively. A similar 
expression holds for the frequency response at port $B$. 
For the circuit in Figure \ref{fig:setup}, it is $N=M=4$. The resonance 
frequencies are the zeros of the denominator in Eq.(\ref{eq:responseA}).

The system's frequency response can be determined by either performing 
measurements in the frequency or time domain. Time domain measurements yield 
more accurate results compared to frequency domain measurements. For a time domain 
measurement the system is excited by a short voltage pulse. The resulting response 
of the input current at a port is the impulse response of the system for that port. 
The Fourier transform of the impulse response is equal to the system's frequency 
response. The frequency response depends on the measurement configuration, i.e. 
the port where the impulse response is measured and the relative amplitude of the 
excitation pulse at ports $A$ and $B$. In Eq.(\ref{eq:responseA}) only the 
roots of the numerator are dependent on the measurement configuration. The roots 
of the denominator, i.e. the eigenvalues, are independent of different measurement 
configurations.

In a practical set-up it is not possible to generate infinitely short pulses. 
However, the pulse width $t_p$ must be small compared to the response time of the 
system. A valid impulse response must obey
\begin{equation}
  t_p \ll \frac{1}{\max(|\omega _i|)}\,\,,
\label{eq:pulseWidth}
\end{equation}
with $\max(|\omega _i|)$ the eigenvalue with the largest modulus.

For this experiment both ports $A$ and $B$ where excited by the same voltage 
pulse. The input current into port $A$ was measured. Applying a Fourier transform 
to the measured impulse response current the frequency response of the system is 
obtained. By fitting Eq.(\ref{eq:responseA}) to the measured data the 
coefficients $a_i$ and $b_i$ can be determined and the resulting eigenvalues of 
the system calculated.

\begin{figure}[h]
  \includegraphics*[scale=0.34]{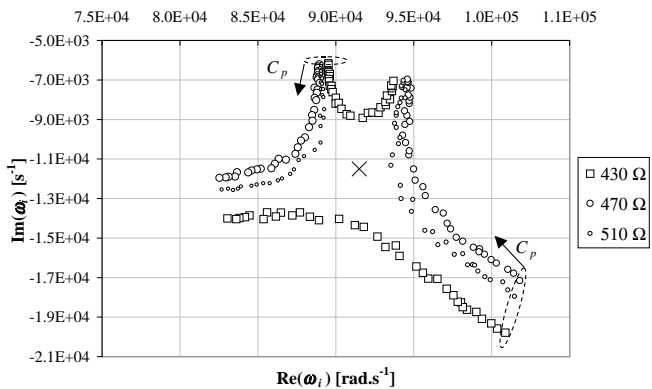}
  \caption{Measured system eigenvalues.}
  \label{fig:eigenPlot}
\end{figure}

The measured eigenvalues for different values of $R_p$ and $C_p$ are plotted on 
the complex energy plane in Figure \ref{fig:eigenPlot}. Only two of the four 
eigenvalues of the system are plotted for each $R_p$ and $C_p$ value pair. The 
other two eigenvalues are mirrored with respect to the imaginary axis. Three 
resistance values were used, $R_p = 430$, $R_p = 470$ and $R_p = 510$. For each 
resistance value the capacitance was changed from 
$57.0\,\mathrm{nF}$ to $72.0\,\mathrm{nF}$ in steps of $0.22\,\mathrm{nF}$. The 
resulting locus of eigenvalues repel from a point in the energy plane, which is 
the singular point of the system. The direction of each locus for an increasing 
value of $C_p$ is indicated in Figure \ref{fig:eigenPlot}. We denote the value 
of the frequency at the singular point by $S_1$ and find 
$S_1 \approx 92000-i11500$ as indicated by a cross in Figure \ref{fig:eigenPlot}. 
A second singular point $S_2$ (not 
shown) is also measured and is mirrored with respect to the imaginary axis of the 
complex energy plane, i.e. $i S_1 = (i S_2)^*$. The measurement error is less 
than $6\,\%$.


In order to measure the system phase, the system has to be excited with a 
steady-state sinusoidal signal. A real sinusoidal signal always consists of two 
frequency components $\omega_1$ and $\omega_2$ for which 
$i \omega_1 = (i \omega_2)^*$. Hence, to 
measure the system phase at the singular point, the system is excited with two 
complex frequencies equal to the two singular point values $S_1$ and $S_2$. 
Furthermore, the circuit values of $R_p$ and $C_p$ are set to obtain eigenvalues 
as close as possible to the two singular points. The excitation signals at port 
$A$ and $B$ have the form:
\begin{equation}
  v(t) = C\sin(\omega_0 t + \phi_v)e^{-\gamma_0 t}\,\,,
  \label{eq:exciteComplex}
\end{equation}
with $\omega_0=|\Re(S_1)|=|\Re(S_2)|$ and $\gamma_0=-\Im(S_1)=-\Im(S_2)$. The 
phase of the signal is $\phi_v$ and the amplitude $C$. For the measured system 
the value of $\gamma_0$ is positive ($\gamma \approx 11500\,s^{-1}$). This 
implies a decaying sinusoidal signal.

The aim of the experiment is to measure the steady-state phase difference 
$\Delta\phi_i$ between the two input currents $i_A$ and $i_B$. This is achieved 
by measuring the steady-state current response at the two ports simultaneously. 
The system is linear and the current response will have the same form as the 
excitation voltage in Eq.(\ref{eq:exciteComplex}). The phase of the two input 
currents, $\phi_{iA}$ and $\phi_{iB}$, can be extracted by fitting 
Eq.(\ref{eq:exciteComplex}) to each of the measured current responses. 
The amplitude 
and phase are the fitting parameters. The phase difference is then simply 
$\Delta\phi_i=\phi_{iA}-\phi_{iB}$.

However, the signal in Eq.(\ref{eq:exciteComplex}) is not a power signal, 
i.e.~it does not have a finite power content. It is therefore not possible in
practice to generate such a signal. To perform the phase measurement the excitation 
is switched on at $t = t_0$. For $t < t_0$ the signal is zero. The on-set of the 
signal at $t = t_0$ introduces a decaying transient response. A reliable 
measurement of the steady-state phase can only be performed at 
$t-t_0 \ll \gamma_0^{-1}$, where the transient response has decayed and can be 
neglected compared to the steady-state response.

To generate signals as described by Eq.(\ref{eq:exciteComplex}) the circuit 
values of the circuit in Figure \ref{fig:setup} were so chosen that singular 
points and eigenvalues are obtained
within the audible frequency range. This enables the excitation signals to be 
generated with a standard computer sound card. LabVIEW is again used 
to drive the sound card of the PC and generate any arbitrary signal within the 
audible frequency range ($20\,\mathrm{Hz}$ to $20\,\mathrm{kHz}$). The stereo 
output of a computer sound card can be used to generate two different signals 
simultaneously.

\begin{figure}[h]
  \includegraphics*[scale=0.34]{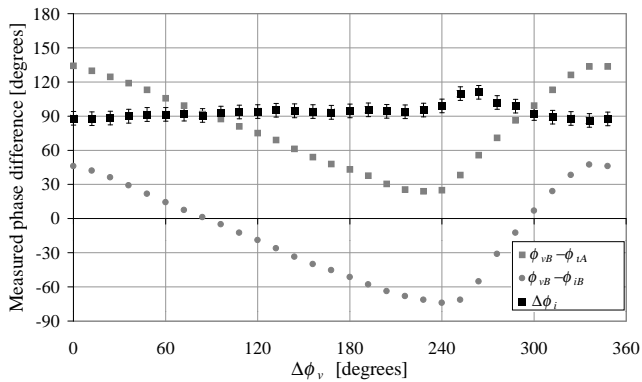}
  \caption{Measured system phase.}
  \label{fig:phaseResult}
\end{figure}

Finally, the steady-state phase difference $\Delta\phi_i$ between the input 
currents was measured while exciting both ports. The phase difference 
$\Delta\phi_v$ between the two excitation signals was varied from 0 to $2\pi$ 
radians. In Figure \ref{fig:phaseResult} the measured current phase difference 
$\Delta\phi_i$ is plotted as a function of the excitation phase difference 
$\Delta\phi_v$. Also shown is the phase difference between the two input currents, 
$i_A$ and $i_B$, and the excitation voltage $v_B$ at port B. The measured phase 
difference $\Delta\phi_i$ is largely independent of the excitation phase difference 
$\Delta\phi_v$ with a value of approximately $\frac{\pi}{2}$ and a measurement 
error of $5\,\%$.

As far as the results are concerned we emphasize in particular two points:
(i) the similarity of Fig.\ref{fig:eigenPlot} with the corresponding figure for
instance in \cite{Latinne} or \cite{DemboEP} is striking; here it refers 
to electronic circuits, in the quoted papers to atomic physics or microwave 
cavities, respectively; (ii) the measured phase difference $\pi /2$ which is independent
of the phase of the excitation, underlines once again the universal and robust
chiral character of the state vector at the EP. 
We mention, that owing to the non-symmetric nature 
of the matrix considered (see Eq.(\ref{mat})) there is in fact a slight 
deviation \cite{he03}. For the parameters chosen the precise quotient of the
amplitudes actually is $-0.08+i 1.99\ldots $ and not $i$ as in Eq.(\ref{restr}). 
In other words, there is a slight deviation from the circular toward an 
elliptic polarization. Yet an orientation prevails.

In this, like in all other experiments where EPs have been observed, dissipation
is playing a foremost role. In fact, in quantum mechanics or optics, only absorption
enables access in the laboratory to the singularity in the complex plane. And even
in the experiment described in the present paper, where the underlying matrices
are no longer symmetric, dissipation is crucial to give rise to an EP. The decay 
of the states provides a direction of the time axis. In classical and quantum systems
absorption -- and decoherence in the latter case \cite{legzu} -- 
occurs due to the interaction with the environment, the
presence of the open channels is effectively described by complex interaction
parameters \cite{he99,weid} or, as in the present case, explicitly by Ohmic resisters.

In turn, the wave function at the EP is usually chiral in character as has been
discussed in \cite{heha} and confirmed experimentally the first time in 
\cite{DemboCh}. We see here an intrinsic connection -- provided by the mathematics
of an EP -- between the direction of time and a given direction in space. In fact,
the present experiment can serve to define a direction in one dimensional space.
At the EP, it is the one oscillator whose phase is leading while the phase of the
other is lagging. These roles are fixed unequivocally for a given parameter set,
unrelated to space. Viewing the oscillators placed along a straight line 
an orientation is defined
along this line. This is not trivial and is caused by the mechanism at an EP: one,
and only one-and-the-same oscillator has the leading phase, and this feature is 
cast in stone by a system that has no {\it a priory} spatial orientation. It is
dissipation that brings about spatial orientation.

This does not mean, at this stage, that a classical set-up can provide a distinction
between left and right. One is tempted to find a truly three-dimensional setting
where one EP might do just that. 

The authors ackowledge financial support by the South African National Research Foundation.
They are grateful to H.B.~Geyer for a critical reading of the manuscript.

\end{document}